\documentclass[aps,prl,preprint,superscriptaddress]{revtex4}

\usepackage{graphicx}
\usepackage{dcolumn}
\usepackage{bm}
\usepackage{color} 

\begin{document}


\title{Unconventional mass enhancement around the Dirac nodal loop in ZrSiS}

\author{S.~Pezzini}
\affiliation{High Field Magnet Laboratory (HFML-EMFL), Radboud University, Toernooiveld 7, Nijmegen 6525 ED, Netherlands.}
\affiliation{Radboud University, Institute for Molecules and Materials, Nijmegen 6525 AJ, Netherlands.}

\author{M.~R.~van~Delft}
\affiliation{High Field Magnet Laboratory (HFML-EMFL), Radboud University, Toernooiveld 7, Nijmegen 6525 ED, Netherlands.}
\affiliation{Radboud University, Institute for Molecules and Materials, Nijmegen 6525 AJ, Netherlands.}

\author{L.~Schoop}
\affiliation{Max Planck Institute for Solid State Research, Heisenbergstr. 1, 70569 Stuttgart, Germany.}

\author{B.~Lotsch}
\affiliation{Max Planck Institute for Solid State Research, Heisenbergstr. 1, 70569 Stuttgart, Germany.}

\author{A.~Carrington}
\affiliation{H. H. Wills Physics Laboratory, University of Bristol, Tyndall Avenue, Bristol BS8 1TL, UK.}

\author{M.~I.~Katsnelson}
\affiliation{Radboud University, Institute for Molecules and Materials, Nijmegen 6525 AJ, Netherlands.}

\author{N.~E.~Hussey}
\email{n.e.Hussey@science.ru.nl}
\affiliation{High Field Magnet Laboratory (HFML-EMFL), Radboud University, Toernooiveld 7, Nijmegen 6525 ED, Netherlands.}
\affiliation{Radboud University, Institute for Molecules and Materials, Nijmegen 6525 AJ, Netherlands.}

\author{S.~Wiedmann}
\email{s.wiedmann@science.ru.nl}
\affiliation{High Field Magnet Laboratory (HFML-EMFL), Radboud University, Toernooiveld 7, Nijmegen 6525 ED, Netherlands.}
\affiliation{Radboud University, Institute for Molecules and Materials, Nijmegen 6525 AJ, Netherlands.}

\date{\today}

\begin{abstract}
\textbf{The topological properties of fermions arise from their low-energy Dirac-like band dispersion 
and associated chiralities. Initially confined to points, extensions of the Dirac dispersion 
to lines and even loops have now been uncovered and semimetals hosting such features have been 
identified. However, experimental evidence for the enhanced correlation effects predicted to 
occur in these topological semimetals has been lacking. Here, we report a quantum oscillation 
study of the nodal loop semimetal ZrSiS in high magnetic fields that reveals significant 
enhancement in the effective mass of the quasiparticles residing near the nodal loop. Above 
a threshold field, magnetic breakdown occurs across gaps in the loop structure with orbits that 
enclose different windings around its vertices, each winding accompanied by an additional $\pi$ 
Berry phase. The amplitudes of these breakdown orbits exhibit an anomalous temperature dependence. 
These findings demonstrate the emergence of novel, correlation-driven physics in ZrSiS 
associated with the Dirac-like quasiparticles.}
\end{abstract}


\maketitle


Condensed-matter systems with an energy-momentum dispersion topologically distinct with 
respect to (i.e. not adiabatically deformable into) the one of standard metals, semiconductors 
or insulators, represent a new exciting frontier in physics \cite{Volovik2007,Kane2005,Bernevig2006,Moore2007,Konig2007}. 
These materials host, in their bulk and/or surfaces, low-energy excitations mimicking relativistic particles, 
which opens new possibilities for the simulation of long-sought phenomena of high-energy physics, 
as well as for the realization of novel (quantum) information schemes. Graphene, topological 
insulators (TIs) and, more recently, Dirac and Weyl semimetals (WSMs) have been pioneering 
platforms for this research area \cite{Young2012,Young2015}.

The advent of nodal-line semi-metals (NLSMs) \cite{Weng2015,Wu2016,Bian2016,Neupane2016,Schoop2016}, 
i.e. systems in which the conduction and valence bands cross each other along a closed trajectory 
(line or loop) inside the Brillouin zone, has opened up a new arena for the exploration of 
topological aspects of condensed matter that may be distinct from those associated with TIs 
or WSMs \cite{Burkov2016,Fang2016}. It is claimed, for example, that due to the vanishing 
density of states near the Fermi level $\epsilon_F$, screening of the Coulomb interaction 
may become weaker in NLSMs than in conventional metals and remain long-ranged \cite{Huh2016}. 
This, coupled with their metallic nature, could make NLSMs more susceptible to various types 
of order (e.g. superconductivity, magnetism or charge order), both in the bulk \cite{Roy2016} 
and on the surface \cite{Liu2016}. 

While the degeneracy points in WSMs are robust against \textit{any} perturbation that preserves 
translational symmetry, nodal lines or nodal loops require additional crystalline 
symmetries, such as mirror or non-symmorphic symmetry, to protect the extended line-like 
touching between the conduction and valence bands \cite{Young2015,Fang2016}. To date, 
only a few candidate NLSMs have been identified experimentally. In non-centrosymmetric 
PbTaSe$_2$, strong spin-orbit coupling leads to the creation of nodal rings which have 
been confirmed by angle-resolved photoemission spectroscopy (ARPES) \cite{Bian2016} while 
in PtSn$_4$, ARPES has revealed a Dirac nodal arc structure that has been attributed 
to surface states \cite{Wu2016}. In both systems, however, the topological elements of 
the electronic structure coexist with several other bands that cross the Fermi level $\epsilon_F$. 

The third experimental class to have been identified thus far is ZrSiX (X = S, Se, Te). 
The first of these, ZrSiS, has a number of distinct features. Firstly, it contains a 
diamond-shaped Fermi surface (FS) close to a line of Dirac nodes that in contrast 
to PbTaSe$_2$ and PtSn$_4$, is the \textit{only} band feature near $\epsilon_F$. Thus, the physical 
behavior of ZrSiS, be it bulk- or surface-derived, is governed essentially by the electronic 
states in close proximity to the nodes. Secondly, the linear dispersion of these bands extends, 
in some regions of the Brillouin zone, over an energy range (2 eV) that is much larger than is 
found in other Dirac-like compounds. The range in which all bands are linearly dispersing 
is $\sim$ 0.5 eV. Spin-orbit coupling introduces a small gap ($\sim$ 0.02 eV) in the Dirac spectrum 
(as illustrated by the energy contour plot shown in Fig. S1 of the Supplementary Information).

ARPES studies have confirmed the existence of the diamond-shaped FS in ZrSiS, within the 
($k_x$, $k_y$) plane, in addition to a Dirac-like dispersion that extends over an energy 
range exceeding 1 eV \cite{Neupane2016,Schoop2016}. Previous quantum oscillation studies 
have also reported evidence for pockets associated with the Dirac nodal loop in ZrSiS 
\cite{Wang2016,Ali2016,Hu2016,Singha2016}, ZrSiSe and ZrSiTe \cite{Hu2016_2,Topp2016}. In 
each case, the topological character of these loops was inferred from the observation of 
a phase shift in the quantum oscillations associated with the Berry phase of the orbit.

While the topological nature of NLSMs appears to be established, evidence for enhanced 
many-body effects, e.g. as a precursor to any new broken symmetry phases, has yet to emerge. 
Here, we report a magnetoresistance study of high-quality ZrSiS single crystals up to 33 Tesla. 
By extending the field range of earlier studies, we are able to resolve many new Shubnikov-de 
Haas (SdH) oscillation frequencies at all temperatures below 60 K, including some very high 
frequency oscillations that arise due to magnetic breakdown across gaps along the nodal loop. 
The effective masses associated with these new frequencies are found to be significantly 
enhanced over conventional band-structure estimates. Moreover, the oscillation amplitude 
for the breakdown orbits exhibits an anomalous temperature dependence reminiscent of that 
recently found in the candidate topological Kondo insulator SmB$_6$ \cite{Tan2015}. Collectively, 
these results provide hints that ZrSiS lies close to a quantum phase transition, possibly to 
some form of density wave order, and is thus an ideal material platform on which to explore 
novel correlation effects in topological matter.

\section{Dirac nodal loop}

\begin{figure}[ht]
\centering
\includegraphics[width=0.75\textwidth]{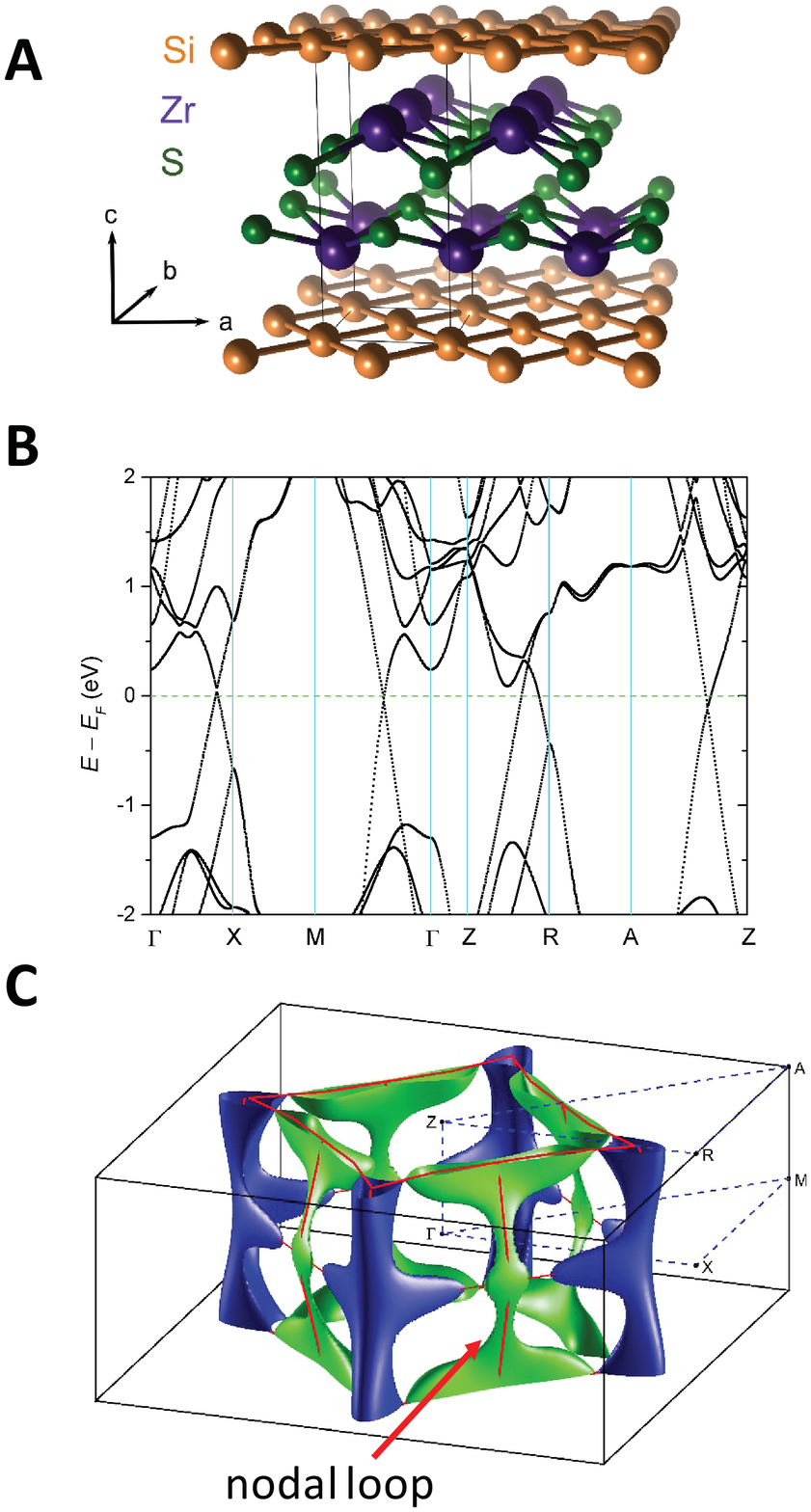}
\caption{}
\label{fig1}
\end{figure}

The crystal structure of ZrSiS (tetragonal space group \textit{P4/nmm}) is displayed in Fig.~\ref{fig1}a. 
ZrSiS has the PbFCl-type structure (like LiFeAs) \cite{Klein1964,Xu2015} with layers of Zr and S 
that are sandwiched between Si square nets extending in the ab-plane. The electronic band structure 
of bulk ZrSiS is shown in Fig.~\ref{fig1}b. The most dominant and noteworthy feature of the 
electronic structure is the series of linearly-dispersing bands that cross very close to $\epsilon_F$, 
giving rise to a nodal loop whose location is indicated in the corresponding Fermi surface plot 
shown in Fig.~\ref{fig1}c. The final assembly of loops gives rise to a diamond-shaped Fermi surface 
within the ($k_x$, $k_y$) plane that is quasi-two-dimensional yet strongly dispersive along $k_z$. 
Significantly, as indicated in Fig.~\ref{fig1}b, the more parabolic bands (mostly originating from 
the sulfur states) are located far from the Fermi level, with the result that the Fermi surface 
depicted in Fig.~\ref{fig1}c is composed uniquely from the almost-linearly dispersing bands.

\section{Oscillatory magnetoresistance}

Figure~\ref{fig2}a shows a series of magnetoresistance (MR) sweeps up to 33 T carried out on a 
ZrSiS single crystal at regular temperature intervals between 1.5 K and 60 K with the magnetic 
field applied perpendicular to the $ab$-plane. Very similar results obtained on a second crystal 
are presented in Fig. S2 of the Supplementary Information. As reported previously 
\cite{Wang2016,Ali2016,Singha2016,Lv2016}, the MR is extremely large, reaching values of order 
10$^4$ \% in 33 T for $\textbf{B}\parallel c$, and 7 $\times$ 10$^5$ \% with the field oriented at a polar 
angle $\theta = 45^{\circ}$. Superimposed on top of the MR background are multiple SdH oscillations, 
some of which are very fast, as emphasized in the blow-up of the highest field data plotted in the 
right panel of Fig.~\ref{fig2}a.
 
A fast Fourier transform (FFT) of the lowest temperature sweep in Fig.~\ref{fig2}a is shown in 
Fig.~\ref{fig2}b. The oscillations can be readily split into two groups; a low frequency ($F <$ 1 kT) 
and a high frequency ($F > $7.5 kT) group. The low frequency spectrum is dominated by two peaks. 
The one at 240 $\pm$ 5 T has been observed by a number of different groups \cite{Ali2016,Hu2016,Singha2016} 
and corresponds to the $\alpha$ hole pocket (the 'petal'), highlighted in blue in the left panel of 
Fig.~\ref{fig2}e, that is located at the vertex of the diamond. The second peak with a frequency 
of 600 $\pm$ 10 T has not been reported previously. It is consistent (albeit 20 \% higher) than band 
structure estimates for the elongated $\beta$ electron pocket (the 'dog-bone') that runs parallel 
to the top rung of the nodal loop (see Fig.~\ref{fig1}c). 

The higher frequency spectrum comprises a series of peaks ranging from 7.5 kT to around 11 kT. 
These peaks do not correspond to any closed contour of the Fermi surface of ZrSiS shown in 
Fig.~\ref{fig1}c but rather likely originate from magnetic breakdown orbits that encircle the 
diamond. This is confirmed by a number of observations. Firstly, as indicated by the dashed 
lines in the right panel of Fig.~\ref{fig2}c, the lower set of five peaks are separated by 
units of 240 T, i.e. by the frequency of the petal orbit. The magnitude of the peaks corresponds 
to a series of breakdown orbits (labelled $A \pm n\alpha$, where $n$ is an integer) that follow 
the inner surface of the dog-bones with each successive peak in the FFT spectrum incorporating 
one additional petal. A similar assignment can be made for the second set of peaks (labelled $B \pm n\alpha$) 
that trace the outer surface of the dog-bones. Fig.~\ref{fig2}e illustrates some of the 
possible breakdown combinations. 

The second telling observation is the rapid suppression of the high frequency oscillations by 
a small tilt of the magnetic field away from the $c$-axis. Fig.~\ref{fig2}c contains a set of 
MR sweeps obtained at different polar angles , while Fig.~\ref{fig2}d shows the corresponding 
high frequency part of the FFT spectrum for increasing $\theta$. The oscillations (and the 
amplitude of the corresponding peaks) are found to be completely suppressed once $\theta > \theta_c$, 
with $2.2^{\circ} < \theta_c < 3.5^{\circ}$. This sudden collapse of the fast oscillations with 
small tilt angle is another clear indication of magnetic breakdown. As illustrated in Fig.~\ref{fig2}f, 
some of the breakdown gaps must increase in size as the field is tilted. Since the tunneling 
probability drops exponentially with gap magnitude (or equivalently, the distance in $k$-space 
between adjacent elements of the breakdown orbit), the oscillation amplitude is then rapidly 
suppressed. A calculation of the attenuation in oscillation amplitude, plotted as a dashed line 
in Fig.~\ref{fig2}g, confirms this (details of the calculation itself can be found in the 
Supplementary Information). 

\section{Topological character}
Having established that the different elements of the Fermi surface are tied to the 
nodal loop, we now turn to examine its topological character. According to the band 
structure calculations, the petal is located at the apex of the diamond and thus 
encircles a Dirac cone (see Fig. S1 of the Supplementary Information for a visualization 
of this). As a consequence, electrons performing any closed circuit including a petal 
should acquire a $pi$ Berry phase \cite{Ali2016}. Given that the breakdown orbits of 
the type $A + n\alpha$ and $B + n\beta$ are closed trajectories that differ by an 
integer number of petals (see Fig.~\ref{fig2}e), one expects there to be a relative 
phase shift of $\pi$ between orbits with $n$ even or odd. This phase shift should then 
be visible in the quantum oscillation trace. Figure~\ref{fig3} shows one such trace 
(black line, after subtraction of a slowly-varying background), together with two 
different simulations for the oscillating component of the resistance (colored lines). 
These simulated curves consist of a sum of cosine terms, with amplitudes and frequencies 
determined from the magnetic breakdown part of the FFT spectrum (the sum is limited to 
those components which are most clearly resolved, see Fig.~\ref{fig2}b). The top panel 
simulation includes a $\pi$ phase for the breakdown orbits corresponding to closed loops 
around an odd number of petals ($A+\alpha, A+3\alpha$,$\ldots$), while the bottom panel 
simulation has no phase factor. The comparison between Figures~\ref{fig3}a and ~\ref{fig3}b s
hows that the Berry phase must be taken into account in order to correctly describe 
the data, thus confirming the topological nature of ZrSiS. Moreover, magnetic breakdown 
orbits that selectively enclose band touching points (i.e. singularities of the Berry 
curvature) are shown here, for the first time, to provide a novel probe of topological systems.

These collective results, together with the more detailed angular dependent study of the 
low frequency oscillations illustrated in Fig. S5 of the Supplementary Information, firmly 
establish the topological nodal loop structure in ZrSiS that was predicted in earlier 
band structure calculations \cite{Schoop2016,Xu2015}. Indeed, for $\textbf{B}\parallel c$, only 
minor corrections to the dog-bone orbit were required to get a consistent match with \textit{all} 
of the peaks identified in the FFT. This is the first main finding of our study. In the 
following, we turn our attention to the quasiparticle masses extracted from these measurements 
and arguably the most surprising finding, namely the anomalous temperature and field 
dependences of the SdH oscillation amplitude.

\section{Mass enhancement along the nodal line} 

Figure~\ref{fig4}a shows a series of raw FFT spectra 
for both the low (left panel) and high (right panel) frequency ranges obtained from the 
full field sweeps performed at different temperatures between 1.5 and 60 K. By restricting 
the field range however, the temperature evolution of the oscillation amplitudes was found 
to vary, implying that the effective masses of the quasiparticles performing each orbit 
were actually field-dependent. The masses obtained from such spectra are shown in Fig.~\ref{fig4}b 
for the petal and dog-bone orbits (see Fig. S6 of the Supplementary Information for the 
corresponding temperature plots). For the petal orbit, $m^*$ varies only slightly with field with 
an average value $m^* = 0.21 \pm 0.03~m_e$, in reasonable agreement with previous low-field 
measurements \cite{Wang2016,Ali2016,Singha2016} and band structure estimates of $m^* = 0.16 \pm 0.02~m_e$. 

For the dog-bone orbit, however, the situation is strikingly different. Firstly, $m^*$ is 
found to have a value larger than 1.0 $m_e$ for all field ranges. This is significantly 
enhanced with respect to the band-derived value of 0.55 $m_e$. Secondly, $m^*$ is found 
to become heavier with increasing field strength. At the highest field range, the 
effective mass of the dog-bone orbit is enhanced by up to a factor of 3. Such an 
enhancement is beyond that expected from a conventional electron-phonon interaction 
and implies significant polaronic or correlation effects. Indeed, this is the largest 
mass enhancement ever observed for a Dirac system and is particularly striking given 
the extremely wide (2 eV) band width of the linear dispersion.

\section{Discussion}

According to Roy and Huh \textit{et al.} \cite{Hu2016,Roy2016}, the specific properties 
of the band dispersions near $\epsilon_F$ in a NLSM, in particular the presence of 
the nodal line running parallel to the Fermi surface, mean that the Coulomb interaction 
is only partially screened due to a vanishing density of states (DOS). In graphene, with 
isolated Dirac points, electron-electron interactions are also only partially screened. 
This leads to a modification of the Dirac dispersion in such a way that the $m^*$ is 
found to shrink as $\epsilon_F$ approaches the Dirac point \cite{Elias2011,Yu2013}. In 
ZrSiS, the Fermi line node is not actually pinned to the Fermi level and therefore the 
DOS does not vanish precisely at $\epsilon_F$. Nevertheless, the fact that the largest 
mass enhancement is found for quasiparticles on the dog-bone pocket which runs parallel 
to the nodal loop suggests that residual Coulomb interactions lead to an \textit{enhancement} 
in $m^*$ rather than a reduction. Smaller, but still finite, effects are also expected 
for the petal orbit. Significantly, ARPES sees no renormalization of the band 
dispersion \cite{Neupane2016,Schoop2016}, suggesting that the renormalization only 
occurs very close to $\epsilon_F$ and is thus unobservable by ARPES.

This field-induced enhancement of $m^*$ for the $\beta$ pocket is reminiscent of what is observed 
in certain correlated electron systems such as YbRh$_2$Si$_2$ \cite{Gegenwart2002} or 
CeCoIn$_5$ \cite{Paglione2003} that can be tuned to a quantum critical point (QCP) by 
a magnetic field; the approach to the QCP is reflected in a divergence of $m^*$ due to a 
dressing of the quasiparticles by the quantum fluctuations associated with the adjacent 
ordered state. In NLSMs, theoretical predictions for continuous quantum phase transitions 
into various ordered states have emerged, both in the bulk \cite{Roy2016} and on the 
surface \cite{Liu2016}, provided onsite or nearest-neighbor interactions are sufficiently 
strong. In ZrSiS, the additional mass enhancement in field possibly arises from the 
raising (through Zeeman splitting) of one of the spin sub-systems towards the Fermi 
level, which in turn (due to the vanishing DOS) leads to enhanced correlation effects. 
In this regard, it would certainly be interesting to explore the evolution of this 
enhancement to even higher fields in due course.

The final, novel finding from our study is the departure from the canonical Lifshitz-Kosevich 
(LK) form of the $T$-dependence of the oscillation amplitudes specifically for the 
breakdown orbits, representative plots of which are shown in Figure~\ref{fig4}c. While 
either the low-$T$ or high-$T$ data can be fitted approximately using the standard LK 
form \cite{Shoenberg1984}, there is a clear discontinuity in all cases at a temperature $T_0$ $\sim$ 8 K. 
The corresponding effective mass extracted from the LK fit above $T_0$ is comparable to 
that obtained for the $\alpha$ pocket, while below $T_0$, it increases by approximately 
one order of magnitude. We are not aware of any physical origin for such a transition, 
particularly given that there is no concomitant change in the Fermi surface topology 
(i.e in the oscillation frequencies) with temperature. Phenomenologically, this type 
of behavior can be attributed to a boson mode with an energy of the order of $T_0$ that 
strongly renormalizes the electron effective mass only within a nonadiabatic 
layer $|\epsilon - \epsilon_F| < T_0$, similar to the case of electron-phonon 
interaction \cite{Schrieffer1983}. More specific assumption on the origin of this 
boson mode will be presented below.

The oscillation amplitude for each breakdown orbit depends not only on the effective mass 
associated with the orbit, but also on the probability to tunnel across the breakdown gap. 
Thus with increasing temperature, the magnetic breakdown may become thermally-assisted, 
leading to an enhancement in the tunneling probability and an upward deviation from the 
LK form for the oscillation amplitude. Within such a scenario, however, one might expect 
the deviation to be gradual \cite{Kishigi1995} and not sharp as observed in ZrSiS, unless 
of course, the breakdown gap itself also shrinks as the temperature is increased. It is 
also worth noting here that no deviations from the LK formula have been reported in 
other breakdown systems such as organic salts or elements such as Zn.

A recent theoretical study of quantum oscillatory phenomena in \textit{gapped} NLSMs predicts 
an anomalous $T$-dependence in the oscillation amplitude \cite{Pal2016}, though not of 
the form reported here for ZrSiS. There, the amplitude is expected to show a non-monotonic 
dependence, collapsing to zero as $T \Rightarrow$ 0 due to the presence of the hybridization 
gap. In ZrSiS, of course, the small hybridization gap is bridged at high fields by the 
breakdown orbits and thus the non-monotonic behavior will be absent. Nevertheless, the 
sharp upturn in the oscillation amplitude found in ZrSiS does not appear to be consistent 
with such a picture.

The sharpness of the deviation is somewhat reminiscent of the behavior reported recently
\cite{Tan2015} in the candidate topological Kondo insulator SmB$_6$ \cite{Dzero2010}. Despite 
having a bulk insulating state, quantum oscillations (in the magnetic torque) were observed 
below 25 K with multiple frequencies up to 15 kT \cite{Tan2015}. Below 1 K, however a dramatic 
upward enhancement of the oscillation amplitude, by almost one order of magnitude, was seen 
in one of the low frequency orbits ($F$ = 330 T). The origin of such oscillations is still 
a subject of ongoing debate and to date, no corresponding oscillations have been seen in 
the electrical resistivity. 

Inspection of the nodal-loop Fermi surface projection of ZrSiS, shown in the inset of 
Fig.~\ref{fig4}, suggests that the unusual mass enhancement on the breakdown orbits 
could be linked to enhanced, large-$Q$ density wave correlations (spin or charge) across 
near-nested sections of the diamond-shaped Fermi surface. In this case, one can expect a 
logarithmically divergent effective mass at the threshold of instability that in principle, 
can make the renormalization extremely strong \cite{Virosztek1990}. The effective energy scale 
of the relevant large-$Q$ boson would have to be very low, however, in order to account for the 
marked change in the effective mass of the quasiparticles involved in the breakdown orbits 
around $T_0$. The field-enhanced mass found for the $\beta$ pocket, by contrast, could arise 
from small-$Q$ density-wave correlations, presumably across the opposite faces of the dog-bone. 
Such a scenario requires both short- \cite{Roy2016} and long-range \cite{Huh2016} many-body interactions 
to be strongly enhanced in ZrSiS.  

The field of topological semi-metals is only just emerging. The addition of correlation 
effects, as suggested by this present study, opens up new frontiers for the exploration of 
novel exotic states in these systems. Indeed, theorists have already predicted the creation 
of novel correlation-induced phase transitions to ordered states, both in the bulk and on 
the surface. Moreover, the associated quantum phase transition is believed to realize 
entirely new critical universality classes \cite{Liu2016}. The work presented in this 
letter suggest that correlation effects are indeed strong in ZrSiS and thus, with careful 
manipulation of this system, e.g. through doping studies, such novel transitions associated 
with the topological nodal loop, at or close to half-filling, could be induced.  

\vspace{2cm}

\section{Methods} 

\subsection{Sample synthesis and characterization}

Single crystals were grown out of the elements using iodine vapour transport. 
Stoichiometric amounts of the elements and a small amount of iodine were placed 
in a carbon coated quartz tube and heated to 1100$^{\circ}$C with a 100$^{\circ}$C temperature 
gradient for 1 week. The obtained crystals were subsequently annealed at 600$^{\circ}$C 
for a period of 4 weeks. The crystal structure was confirmed with 
single-crystal X-ray and electron diffraction.

\subsection{Magnetotransport measurements}

We studied two samples with parallel-piped shape. Six electrical contacts 
were defined on each sample with silver conductive paste; two of them, uniformly 
covering the two smaller faces of the samples, were employed as current source 
and drain, while the other four acted as voltage probes on different lateral 
sides (equivalent data were collected on the two lateral sides of the samples, 
indicating high homogeneity in terms of charge distribution and current flow). 
The resistance data were acquired in four probe configuration, with a constant 
current excitation of 3 mA, and standard lock-in acquisition at 13 Hz frequency. 
The two samples were attached to a rotating stage and simultaneously characterized 
in a $^4$He VTI system with base temperature 1.5 K. The cryogenic system was 
accommodated in the room-$T$ access bore (32 mm diameter) of a resistive Bitter magnet 
at HFML, with maximum field 33 T. The samples displayed highly symmetric resistance 
signals with respect to the magnetic field direction, hence only data obtained for 
positive magnetic field orientation are shown in the paper.

\subsection{Electronic structure calculations}

Electronic structure calculations were performed using a full-potential linearized 
augmented plane-wave plus local orbital method as implemented in the Wien2K 
package \cite{Blaha2011}. The experimental lattice parameters and internal positions 
were used ($a$ = 3.544 {\AA}, $c$ = 8.055 {\AA}, $Z_{Zr}$ = 0.2725, $Z_S$= 0.6220 \cite{Klein1964}) 
along with the PBE-GGA exchange correlation potential \cite{Perdew1996}. The spin-orbit 
interaction was included using a second variational method \cite{Blaha2011}. The 
calculation was converged with 10$^4$ $k$-points in the full Brillouin zone. For 
rendering of the Fermi surface (Fig.~\ref{fig1}c) the energy eigenvalues were 
evaluated on a much more dense grid of 4 $\times$ 10$^6$ $k$-points in order to 
minimise interpolation errors which become severe as the Fermi level approaches the 
nodal line. In order to accurate calculate the breakdown gap, an even more 
dense mesh with $k$-point spacing smaller than 10$^{-3}$ $a$ was necessary.

\section*{Acknowledgements}
We acknowledge enlightening discussions with Y.-B. Kim and A. McCollam. We also 
acknowledge the support of the HFML-RU/FOM, member of the European Magnetic Field 
Laboratory (EMFL). A portion of this work was supported by the Engineering and 
Physical Sciences Research Council (grant no. EP/K016709/1). 

\section*{Author contributions}
S.W. initiated the project in collaboration with L.S.. S.P., M.v.D. and S.W. 
performed the magnetotransport measurements. L.S. and B.L. synthesized the ZrSiS 
single crystals. A.C. performed the electronic band structure calculations. 
S.P., M.v.D., S.W., A.C., M.I.K. and N.E.H. analysed the data. N.E.H. wrote 
the manuscript with input from all the co-authors.


\section*{Materials and correspondence}
Correspondence and requests for materials should be
addressed to S.W. (s.wiedmann@science.ru.nl) and N.E.H. (n.e.hussey@science.ru.nl).

\clearpage

\begin{figure}
	\centering
	\includegraphics[width=0.45\textwidth]{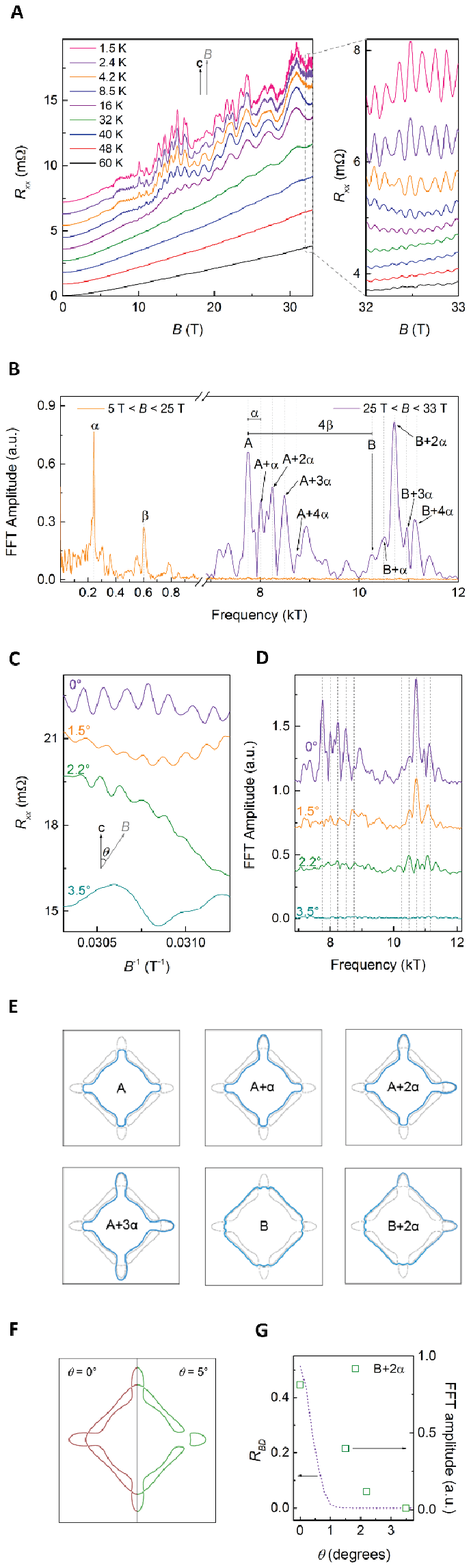}
	\caption{}
	\label{fig2}
\end{figure}

\begin{figure}
	\centering
	\includegraphics[width=\textwidth]{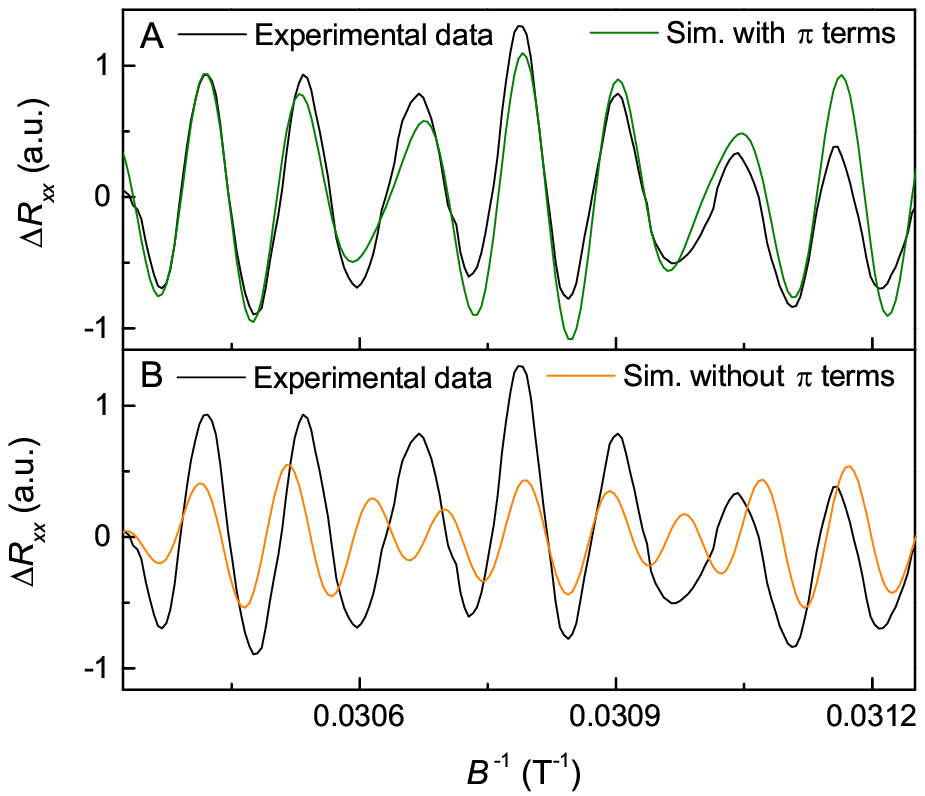}
	\caption{}
	\label{fig3}
\end{figure}

\begin{figure}
	\centering
	\includegraphics[width=0.9\textwidth]{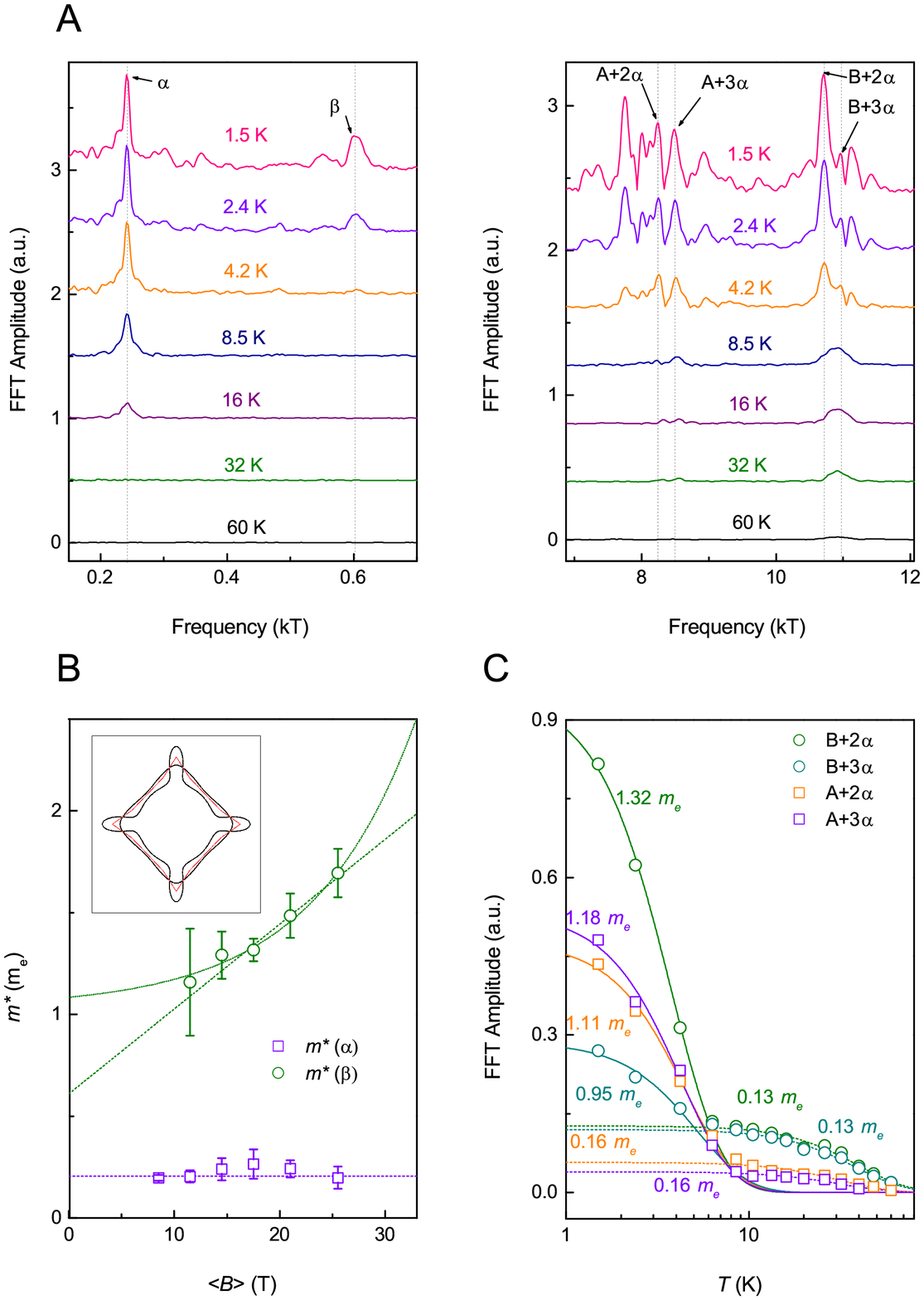}
	\caption{}
	\label{fig4}
\end{figure}

\clearpage

\section{Figure captions}

\textbf{Figure 1: Crystallographic and electronic structure of ZrSiS}. 
(a) Crystal structure of ZrSiS. The Si square net can be seen in blue. (b) Calculated 
bulk band structure with spin-orbit coupling. (c) Resultant three-dimensional 
Fermi surface. The top surface of the Brillouin zone contains the symmetry points 
Z, R and A. The location of the Dirac nodal loop is shown by the dashed black line.\\
\\
\textbf{Figure 2: Shubnikov-de Haas oscillations and breakdown orbits in ZrSiS in a 
perpendicular magnetic field} (a) Left panel: Series of magnetoresistance (MR) 
sweeps up to 33 T with $\textbf{B}\parallel c$ at the temperatures listed (curves are offset 
for clarity). Right panel: Close-up of the high-field part of the sweep, highlighting 
the fast oscillations that are associated with the onset of magnetic breakdown. (b) 
Fast Fourier transform (FFT) of the full MR sweep, having subtracted off a smooth 
polynomial background fit. The spectrum is divided into low a frequency part encompassing 
the frequencies due to the $\alpha$ (petal) pocket ($F$ = 240 T) and $\beta$ (dog-bone) 
pocket ($F$ = 600 T) and a high frequency part that includes the frequencies of the 
breakdown orbits. The magnetic field ranges considered for the two parts are indicated 
in the figure. Label $A$ corresponds to the 'inner' breakdown orbits, $B$ to the 
'outer' orbits as illustrated in Fig. 2(e). (c) Polar angle dependence of the high-field 
MR sweeps as a function of 1/$B$ and (d) corresponding FFT spectra, showing the strong 
suppression of the high frequency (breakdown) oscillations with tilt angle. (e) 
In-plane projection of the diamond Fermi surface at $\theta$ = 0$^{\circ}$; the outer 
square represents the first Brillouin zone. Several of the breakdown orbits are indicated, 
their labels refer to the corresponding peaks in the FFT spectrum. (f) In-plane projection 
of the diamond Fermi surface for $\theta$ = 0$^{\circ}$ (left panel) and $\theta$ = 5$^{\circ}$ 
(right panel). In a tilted field, certain breakdown gaps increase in magnitude. 
(g) Attenuation of the oscillation amplitude as a function of $\theta$. The solid 
line is a calculation of the damping term. See the Supplementary Information for more details.\\
\\
\textbf{Figure 3: Evidence for the topological character of the nodal loop.} (a), (b) Black line: 
Oscillatory part of the magnetoresistance trace at 1.5 K ($\textbf{B}\parallel c$), as a function of 
1/$B$, between 32 T and 33 T. The green curve in (a) corresponds to 
$\sum_n a_n$cos($2\pi F_{A,n}/B + n\pi$) + $b_n$cos($2\pi F_{B,n}/B + n\pi$), i.e. a sum of 
cosine functions with amplitudes ($a_n$ and $b_n$) and frequencies ($F_{A,n}$ and $F_{B,n}$) given 
by the main peaks of the FFT in the right part of Figure~\ref{fig2}b, and including a non-trivial Berry 
phase term ($n\pi$) for the breakdown orbits encircling an odd numbers of band-touching vertexes. 
The orange trace in (b) is the same function of the green one in (a), without the $n\pi$ terms.\\
\\
\textbf{Figure 4: Effective mass and anomalous thermal damping of the oscillation amplitude.} (a) 
Fast Fourier transforms of the full magnetoresistance sweeps performed with $\textbf{B}\parallel c$ at 
the different temperatures listed for both the (left panel) low-frequency and (right panel) 
high-frequency oscillations (same magnetic field ranges of Fig~\ref{fig2}b). (b) Effective masses 
for the $\alpha$ and $\beta$ pockets obtained from $T$-dependence of the oscillation amplitudes 
over the different field ranges as indicated. The dashed and dashed-dotted lines are guides to 
the eye, under the assumption that the zero-field mass is unrenormalized or renormalized respectively. 
Inset: location of the nodal line (dashed line) within the in-plane projection of the diamond Fermi 
surface for $\textbf{B}\parallel c$. (c) Mass plots for a number of frequencies corresponding to breakdown orbits. 
The solid and dashed lines are fits to the LK expression below and above $T_0$ = 8 K respectively. 
Again, the quasiparticle masses extracted from each fit are given in the Figure.\\

\end{document}